\documentclass[3p]{elsarticle}

\usepackage{hyperref}

\journal{}

\biboptions{authoryear}

\usepackage{multirow}
\usepackage{graphicx}
\usepackage{subfig}
\usepackage{amsmath}
\usepackage{amssymb}
\usepackage{hyperref}
\usepackage{booktabs,tabularx}
\newcolumntype{x}{>{\centering\arraybackslash}X}
\usepackage{float}
\usepackage[]{algorithm2e}

\usepackage{tikz}
\usetikzlibrary{arrows,decorations.pathmorphing,backgrounds,positioning,fit,petri}
\usetikzlibrary{bayesnet}
\usetikzlibrary{mindmap,trees}

\newcommand{\transpose}{{\mbox{\scriptsize T}}}

\newcommand{\N}{\mathcal{N}}
\newcommand{\E}{\mathbb{E}}
\newcommand{\bs}{\boldsymbol}
\newcommand{\bx}{\textbf{x}}
\newcommand{\bX}{\textbf{X}}

\newcommand{\by}{\textbf{y}}

\newcommand{\bz}{\textbf{z}}

\begin{document}

\begin{frontmatter}

\title{Scaling Bayesian inference of mixed multinomial logit models to very large datasets} 


\author[mymainaddress]{Filipe Rodrigues\corref{mycorrespondingauthor}}
\ead[url]{http://fprodrigues.com}

\cortext[mycorrespondingauthor]{Corresponding author}
\ead{rodr@dtu.dk}

\address[mymainaddress]{Technical University of Denmark (DTU), Bygning 116B, 2800 Kgs. Lyngby, Denmark}

\begin{abstract}
Variational inference methods have been shown to lead to significant improvements in the computational efficiency of approximate Bayesian inference in mixed multinomial logit models when compared to standard Markov-chain Monte Carlo (MCMC) methods without compromising accuracy. 
However, despite their demonstrated efficiency gains, existing methods still suffer from important limitations that prevent them to scale to very large datasets, while providing the flexibility to allow for rich prior distributions and to capture complex posterior distributions. In this paper, we propose an Amortized Variational Inference approach that leverages stochastic backpropagation, automatic differentiation and GPU-accelerated computation, for effectively scaling Bayesian inference in Mixed Multinomial Logit models to very large datasets. Moreover, we show how normalizing flows can be used to increase the flexibility of the variational posterior approximations. Through an extensive simulation study, we empirically show that the proposed approach is able to achieve computational speedups of multiple orders of magnitude over traditional MSLE and MCMC approaches for large datasets without compromising estimation accuracy. 
\end{abstract}


\end{frontmatter}


\section{Introduction}
\label{sec:introduction}

The mixed multinomial logit (MMNL) model \citep{mcfadden2000mixed} currently dominates research and practice when it comes to understanding and predicting individual choice behavior, with tremendous impact in many areas such as policy making for sustainability, energy systems, transportation, healthcare, urban design and marketing. Parameter estimation in the MMNL model is typically performed in a maximum likelihood framework, typically using a maximum simulated likelihood estimation (MSLE) approach \citep{train2009discrete}, as currently implemented by popular discrete choice analysis frameworks and tools: \mbox{e.g.} Biogeme \citep{bierlaire2003biogeme} and PyLogit \citep{brathwaite2018asymmetric}. However, assuming a fully-Bayesian approach to the MMNL and performing Bayesian inference carries several advantages over maximum likelihood estimation, such as the ability of obtaining full posterior distributions over the model parameters (including the the individual-specific taste parameters) through a proper treatment of uncertainty, the ability to handle incomplete data by marginalizing over missing variables, the natural support for online inference for streaming data (\mbox{e.g.} as in \cite{danaf2019online}), the support to automatic utility function specification approaches \citep{rodrigues2019bayesian}, etc.

Unfortunately, performing Bayesian inference in the MMNL model is not trivial and requires the use of approximate inference methods. While Markov-chain Monte Carlo (MCMC) methods are the predominant choice in the MMNL literature for posterior inference \citep{train2009discrete,danaf2019online}, in practice, their application is hampered by their extremely high computational costs (both in terms of time and storage), and the difficulty in assessing convergence and diagnosing the quality of the samples obtained \citep{braun2010variational,blei2017variational,bansal2020bayesian}. Although MCMC methods have the appealing asymptotic property of converging to the true posterior distribution in the limit of infinite computational time, variational inference (VI) \citep{Jordan1999,Wainwright2008} provides a significantly advantageous paradigm. Variational inference methods consider a tractable parametric approximate distribution, whose parameters are optimized in order to make it as close as possible to the true posterior distribution (typically measured in terms of the Kullback-Leibler divergence), thereby transforming Bayesian inference into an optimization problem. In doing so, VI tends to significantly outperform MCMC methods without compromising posterior estimation accuracy \citep{braun2010variational}. \cite{bansal2020bayesian} provide an excellent comparison study of VI methods with MCMC and MSLE for the MMNL model, with particular emphasis on different techniques for approximating the log-sum-exp term in the variational lower bound. However, despite demonstrating significant improvements in computational performance, the use of standard VI methods (\mbox{e.g.} as in \cite{bansal2020bayesian}) still suffers from important limitations such as (i) the scalability to very large datasets due to the linear growth in the number of parameters in the variational distribution and the use of the entire dataset to compute gradients, (ii) the inability to use modern priors due to the reliance on conjugacy, and (iii) the lack of flexibility in the variational distribution to capture highly complex posteriors. 

With these challenges in mind, this paper proposes an Amortized Variational Inference \citep{zhang2018advances} approach for effectively scaling Bayesian inference in Mixed Multinomial Logit models to very large datasets. The core idea is to parameterize the variational distribution over the local random taste parameters with a deep neural network in order to avoid the growth in variational parameters with the number of decision-makers in the dataset. Together with the use stochastic backpropagation \citep{rezende2014stochastic,kingma2013auto}, automatic differentiation \cite{paszke2017automatic} and GPU-accelerated computation, the proposed approach is empirically shown to be able to achieve computational speedups of multiple orders of magnitude over traditional MSLE and MCMC approaches for large datasets without compromising estimation accuracy. By relying on efficient Monte Carlo gradient estimation techniques \citep{mohamed2019monte}, the proposed approach does not require the use of conjugate priors. Moreover, in order to obtain very flexible variational distributions, we propose the use of normalizing flows \citep{rezende2015variational,papamakarios2019normalizing}, in which a simple base distribution (\mbox{e.g.} Gaussian) is transformed by a series of consecutive bijective differentiable transformations whose parameters are estimated as part of the optimization of the variational lower bound. All these properties are demonstrated empirically using an extensive simulation study. 

The remainder of this paper is organized as follows: Section~\ref{sec:mmnl} presents a full Bayesian formulation of the MMNL model; Section~\ref{sec:approach} describes our proposed Amortized Variational Inference approach and all aspects related to it (\mbox{e.g.} stochastic backpropagation, inference network and normalizing flows). The empirical results from our extensive simulation study are provided in Section~\ref{sec:experiments}, and finally we provide conclusions in Section~\ref{sec:conclusion}. 

\section{Mixed Multinomial Logit Model}
\label{sec:mmnl}

Let us consider a standard discrete choice setup where, on each choice occasion (or menu) $t \in \{1,\dots,T_n\}$, a decision-maker $n \in \{1,\dots,N\}$ derives a random utility $U_{ntj} = V(\textbf{x}_{ntj},\bs\eta_n) + \epsilon_{ntj}$ from each alternative $j$ in the choice set $\mathcal{C}_{nt}$. The systematic (or deterministic) utility term $V(\textbf{x}_{ntj},\bs\eta_n)$ is assumed to be a function of covariates $\textbf{x}_{ntj}$ and a collection of taste parameters $\bs\eta_n$, while $\epsilon_{ntj}$ is a stochastic noise term. Assuming that $\epsilon_{ntj}$ follows a type-I Extreme Value distribution, $\epsilon_{ntj} \sim \mbox{EV}(0,1)$, leads to the standard Multinomial Logit Model (MNL) kernel \citep{mcfadden1973conditional}, according to which the probability of the decision-maker $n$ selecting alternative $j$ is given by
\begin{align}
p(y_{nt} = j|\textbf{x}_{ntj},\bs\eta_n) = \frac{ e^{V(\textbf{x}_{ntj},\bs\eta_n)} }{ \sum_{k \in \mathcal{C}_{nt}} e^{V(\textbf{x}_{ntk},\bs\eta_n)} }.
\label{eq:mnl_kernel}
\end{align}
For the sake of simplicity, we assume the systematic utility function, $V(\textbf{x}_{ntj},\bs\eta_n)$, to be linear-in-parameters. We also consider the general setting under which the tastes $\bs\eta_n$ can be decomposed into a vector of fixed taste parameters $\bs\alpha \in \mathbb{R}^L$ that are shared across decision-makers, and random taste parameters $\bs\beta_n \in \mathbb{R}^K$, which are individual-specific. The systematic utility function is then defined as
\begin{align}
V(\textbf{x}_{ntj},\bs\eta_n) = \bs\alpha^\transpose\textbf{x}_{ntj,F} + \bs\beta_n^\transpose\textbf{x}_{ntj,R}, 
\end{align}
where the decomposition $\textbf{x}_{ntj}^\transpose = (\textbf{x}_{ntj,F}^\transpose, \textbf{x}_{ntj,R}^\transpose)$ is used to distinguish between covariates that pertain to the fixed parameters $\bs\alpha$ and random parameters $\bs\beta_n$, respectively. All vectors are assumed to be column vectors. 

Following the standard mixed multinomial logit (MMNL) model \citep{mcfadden2000mixed} formulation, we assume the random taste parameters $\bs\beta_n$ (individual-specific) to be distributed according to a multivariate normal, \mbox{i.e.} $\bs\beta_n \sim \N(\bs\zeta,\bs\Omega)$, where $\bs\zeta$ is $K$-dimensional mean vector and $\bs\Omega$ is a $K \times K$ covariance matrix. The remaining (global) parameters, $\bs\alpha$, $\bs\zeta$ and $\bs\Omega$, are treated in a fully-Bayesian manner. The fixed taste parameters $\bs\alpha$ and the mean vector $\bs\zeta$ are assumed to follow multivariate normal distributions: $\bs\alpha \sim \N(\bs\lambda_0, \bs\Xi_0)$ and $\bs\zeta \sim \N(\bs\mu_0, \bs\Sigma_0)$. As for the covariance matrix $\bs\Omega$, as recommended by \cite{gelman2006data} and \cite{barnard2000modeling}, we decompose our prior into a scale and a correlation matrix as follows
\begin{align}
\bs\Omega = \mbox{diag}(\bs\tau) \times \bs\Psi \times  \mbox{diag}(\bs\tau),
\end{align}
where $\bs\Psi$ is a correlation matrix and $\bs\tau$ is the vector of coefficient scales. The components of the scale vector $\bs\tau$ are then given a vague half-Cauchy prior, \mbox{e.g.} $\tau_k \sim \mbox{half-Cauchy}(10)$, while for the correlation matrix $\bs\Psi$ we employ a LKJ prior \citep{lewandowski2009generating}, such that $\bs\Psi \sim \mbox{LKJ}(\nu)$, where the hyper-parameter $\nu$ directly controls the amount of correlation that the prior favours. This approach contrasts with the one in \cite{bansal2020bayesian}, where an inverse Wishart is used as a prior for $\bs\Omega$. That choice is motivated primarily by the inverse Wishart prior being conjugate to the multivariate normal, which in turn simplifies the analytical derivations of the Gibbs sampler and variational inference algorithms. However, since in this work we focus on black-box variational inference algorithms \citep{ranganath2014black}, these constrains don't apply, thereby allowing us to employ modern alternatives that provide, among other advantages, a more natural interpretation of the hyper-parameters \citep{lewandowski2009generating}. 

The generative process of the fully Bayesian MMNL model can then be summarised as follows: 
\begin{enumerate}
\item Draw fixed taste parameters $\bs\alpha \sim \N(\bs\lambda_0, \bs\Xi_0)$
\item Draw mean vector $\bs\zeta \sim \N(\bs\mu_0, \bs\Sigma_0)$
\item Draw scales vector $\bs\tau \sim \mbox{half-Cauchy}(\bs\sigma_0)$
\item Draw correlation matrix $\bs\Psi \sim \mbox{LKJ}(\nu)$
\item For each decision-maker $n \in \{1,\dots,N\}$
\begin{enumerate}
\item Draw random taste parameters $\bs\beta_n \sim \N(\bs\zeta,\bs\Omega)$
\item For each choice occasion $t \in \{1,\dots,T_n\}$
\begin{enumerate}
\item Draw observed choice $y_{nt} \sim \mbox{MNL}(\bs\alpha, \bs\beta_n, \bX_{nt})$
\end{enumerate}
\end{enumerate}
\end{enumerate}

According to this generative process, and letting $\bz = \{\bs\alpha,\bs\zeta,\bs\tau,\bs\Psi,\bs\beta_{1:N}\}$ denote the set of all latent variables in the MMNL model, the joint distribution of the model factorises as
\begin{align}
p(\by_{1:N}, \bz) = p(\bs\alpha|\bs\lambda_0, \bs\Xi_0) \, p(\bs\zeta|\bs\mu_0, \bs\Sigma_0) \, p(\bs\tau|\bs\sigma_0) \, p(\bs\Psi|\nu) \prod_{n=1}^N p(\bs\beta_n|\bs\zeta,\bs\Omega) \prod_{t=1}^T p(y_{nt}|\bX_{nt}, \bs\alpha, \bs\beta_n),
\label{eq:joint}
\end{align}
where we introduced the vector notation $\by_n = (y_{n1},\dots,y_{nT})^\transpose$. The goal of Bayesian inference is then to compute the posterior distribution of $\bz$ given a dataset of observed choices. Making use of Bayes' rule, this posterior is given by
\begin{align}
p(\bz|\by_{1:N}) = \frac{p(\by_{1:N}, \bz)}{\int p(\by_{1:N}, \bz) \, d\bz},
\end{align}
which, unfortunately, is intractable to compute due to the high-dimensional integral in the denominator. Therefore, one must resort to approximate inference methods. In the next section, we will propose a flexible, efficient and highly scalable strategy for performing approximate inference in the MMNL model. 

\section{Amortized Variational Inference}
\label{sec:approach}

Following the theory of variational inference, or \emph{variational Bayes} \citep{Jordan1999,Wainwright2008}, we construct an approximation to the true posterior distribution $p(\bz|\by)$ by considering a family of tractable distributions $q_{\bs\phi}(\bz|\by)$, where we dropped the range in $\by_{1:N}$ to keep the notation uncluttered. The inference problem is then reduced to an optimization problem, where the goal is find the parameters $\bs\phi$ of the variational approximation $q_{\bs\phi}(\bz|\by)$ that make it as close as possible to the true posterior. The closeness between the approximate posterior and the true posterior can be measured by the Kullback-Leibler (KL) divergence \citep{MacKay2002}, which is given by
\begin{align}
\mathbb{KL}(q_{\bs\phi}(\bz|\by)||p(\bz|\by)) = \int q_{\bs\phi}(\bz|\by) \log \frac{q_{\bs\phi}(\bz|\by)}{p(\bz|\by)} \, d\bz.
\label{eq:ch2_kl_divergence}
\end{align}
Unfortunately, the KL divergence in (\ref{eq:ch2_kl_divergence}) cannot be minimized directly. However, we can find a function that we can minimize, which is equal to it up to an additive constant, as follows
\begin{align}
\mathbb{KL}(q_{\bs\phi}(\bz|\by)||p(\bz|\by)) &= \mathbb{E}_q \bigg[\log \frac{q_{\bs\phi}(\bz|\by)}{p(\bz|\by)}\bigg]\nonumber\\
&= \E_q [\log q_{\bs\phi}(\bz|\by)] - \E_q [\log p(\bz|\by)]\nonumber\\
&= -(\underbrace{\E_q [\log p(\by,\bz)] - \E_q [\log q_{\bs\phi}(\bz|\by)]}_{\mbox{\scriptsize $\mathcal{L}(q)$}}) + \underbrace{\log p(\by)}_{\mbox{\scriptsize const.}}.
\label{eq:elbo_deriv1}
\end{align}
The $\log p(\by)$ term in (\ref{eq:elbo_deriv1}) does not depend on $\bs\phi$ and thus can be ignored. Minimizing the KL divergence is then equivalent to maximizing $\mathcal{L}(q)$, which is referred to as the evidence lower bound (ELBO).

\subsection{Stochastic Backpropagation}
\label{subsec:sb}

Maximizing the ELBO, $\mathcal{L}(q)$, \mbox{w.r.t.} the parameters $\bs\phi$ requires us to compute gradients $\nabla_{\bs\phi} \mathcal{L}(q)$. While $\nabla_{\bs\phi} \E_q [\log q_{\bs\phi}(\bz|\by)]$ can generally be computed analytically given a tractable choice of approximate distribution (\mbox{e.g.} fully-factorized, or mean-field, approximation \citep{Jordan1999}), computing $\nabla_{\bs\phi} \E_q [\log p(\by,\bz)]$ exactly is infeasible for the MMNL described in Section~\ref{sec:mmnl}, regardless of the choice of $q_{\bs\phi}(\bz|\by)$. In fact, this holds true for a wide majority of models of interest \citep{ranganath2014black}. In the particular case of the multinomial logit, the difficulty stems from the denominator term in (\ref{eq:mnl_kernel}), which requires the computation of an expectation of a log-sum of exponentials (LSE). Although several local approximations and bounds have been proposed in the literature (\mbox{e.g.} \cite{braun2010variational,knowles2011non}) for the expectation of the LSE term (see \cite{bansal2020bayesian} for a detailed comparison of different approaches), here we shall consider a general approach for non-conjugate models using Monte Carlo gradient estimation \citep{mohamed2019monte}. The reason is three-fold: (i) it frees us from conjugacy requirements and allows to easily consider and experiment with different modern priors (\mbox{e.g.} as the LKJ prior for correlation matrices) and model extensions in a probabilistic programming context \citep{tran2017deep,wood2014new}; (ii) the computational overhead is minimal in order to be considered a relevant factor; and (iii) as we shall empirically demonstrate, in combination with automatic differentiation \citep{paszke2017automatic}, amortization \citep{zhang2018advances} and GPU-accelerated computation, this approach leads to excellent scalability properties. 

In essence, our approach consists in considering a flexible family of approximate distributions $q_{\bs\phi}(\bz|\by)$ parameterized by a deep neural network (see Section~\ref{subsec:in} for details). In order to compute gradients of $\mathcal{L}(q)$ \mbox{w.r.t.} all the variational parameters in $\bs\phi$, we rely on \emph{stochastic backpropagation} \citep{rezende2014stochastic} (also known as \emph{stochastic gradient variational Bayes} \citep{kingma2013auto}) - a technique that combines the use of a non-centered reparameterization of the expectation in $\mathcal{L}(q)$ with Monte Carlo approximation. We reparameterize the latent variables in the MMNL model, $\bz$, in terms of a known base distribution and a differentiable transformation. For example, if for a given latent variable $z$ in the model we assume $q_{\bs\phi}(z) = \N(z|\mu,\sigma^2)$, with $\bs\phi = \{\mu,\sigma\}$, we can reparameterize it as
\begin{align}
z \sim \N(z|\mu,\sigma^2) \Leftrightarrow z = \mu + \sigma\epsilon, \quad \epsilon \sim \N(0,1).
\end{align}
We can then compute gradients of an arbitrary function of $z$ (\mbox{e.g.} the ELBO) \mbox{w.r.t.} $\bs\phi$ using a Monte Carlo approximation with draws from the base distribution
\begin{align}
\nabla_{\bs\phi} \E_{q_{\bs\phi}(z)}[f(z)] \Leftrightarrow \E_{\N(\epsilon|0,1)} [ \nabla_{\bs\phi} f(\mu + \sigma\epsilon)].
\label{eq:rep_trick2}
\end{align}
Crucially, notice how the expectation no longer depends on the parameters $\bs\phi$. This means that we can draw samples from $\N(\epsilon|0,1)$ in order to approximate the required gradient, which in turn can be calculated using modern automatic differentiation tools \citep{paszke2017automatic}. For random variables that are not easily reparameterizable, other approaches exist. The discussion is out of scope of this paper, but the interested reader is redirected to \cite{mohamed2019monte}. 

In practice, this type of stochastic gradient approximation is typically accompanied by the use of mini-batching, where small subsamples of the data (mini-batches) are used to approximate the ELBO gradients. This is commonly referred to as Stochastic Variational Inference (SVI), and it has been shown to significantly speed up inference, particularly in large datasets, without compromising estimation accuracy \citep{hoffman2013stochastic}. The combination of these two types of stochasticity, \mbox{i.e.} from the Monte Carlo gradient approximation in (\ref{eq:rep_trick2}) and the stochasticity induced by the use of mini-batches, is also often referred to as Doubly-Stochastic Variational Inference \citep{titsias2014doubly}. Convergence is assumed when the ELBO does not improve for a number of consecutive iterations. 

\subsection{Inference Network}
\label{subsec:in}

\begin{figure}[t!]
\begin{center}
\includegraphics[width=0.8\linewidth,trim=0 0 0 0, clip]{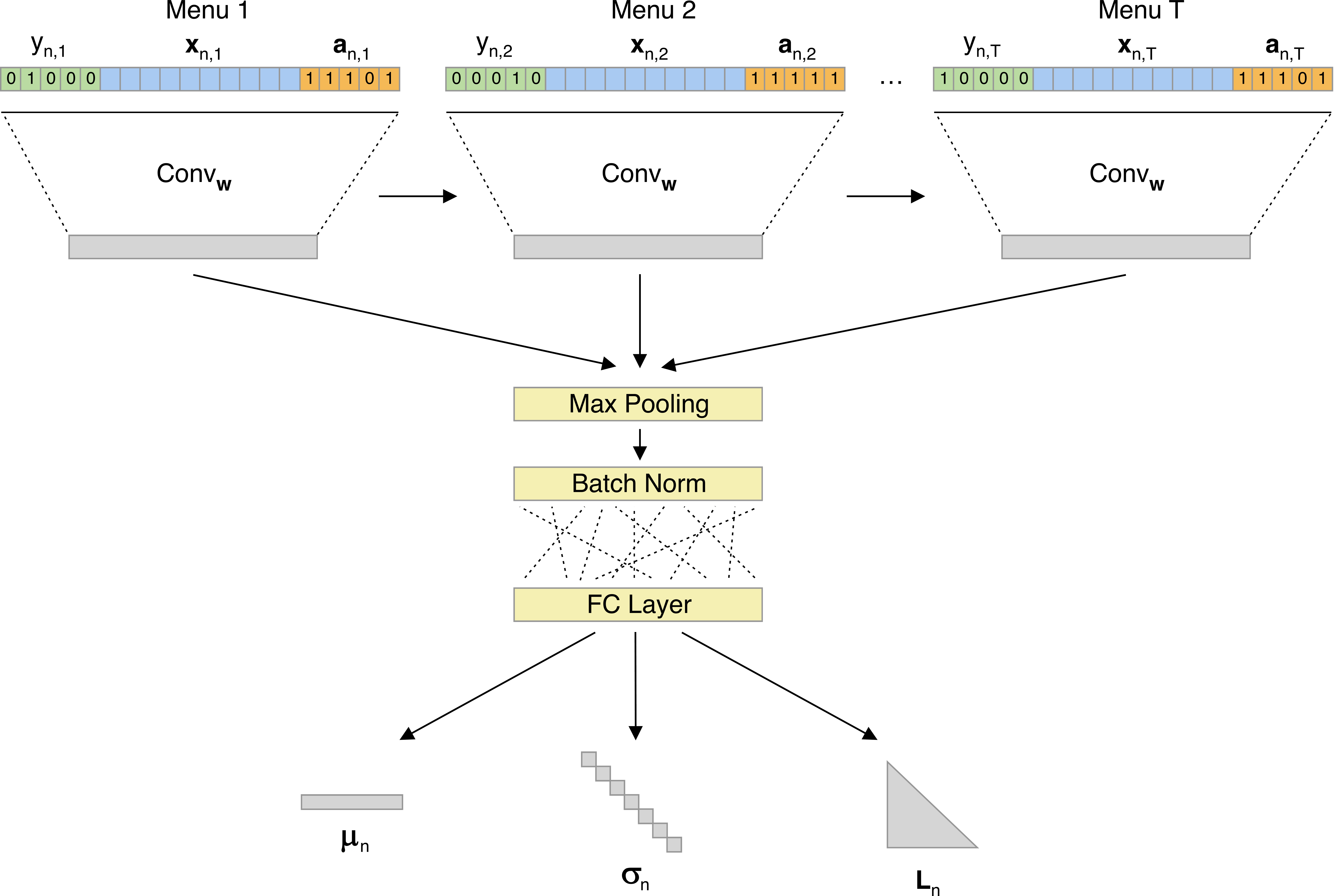}
\caption{Inference Network.}
\label{fig:inference_net}
\end{center}
\end{figure}

So far, the form of the variational distribution $q_{\bs\phi}(\bz|\by)$ hasn't been discussed. A standard approach would be to consider a fully-factorized (mean-field) approximation as proposed in \cite{bansal2020bayesian}, which in this case would take the form
\begin{align}
q_{\bs\phi}(\bz|\by) = q(\bs\alpha) \, q(\bs\zeta) \, q(\bs\tau) \, q(\bs\Psi) \prod_{n=1}^N q(\bs\beta_n),
\end{align}
where we omitted the variational parameters of each term in the factorization in order to simplify the presentation. 
Although for small datasets this can be a good approach, for larger datasets, the fact that the number of local variational parameters grows with the number of decision-makers $N$ (due to the individual taste parameters $\bs\beta_n$) can be problematic. We propose to address this issue using \emph{amortization} \citep{zhang2018advances}. Instead of introducing local variational parameters per decision-maker $n$ by assuming $q(\bs\beta_n) = \N(\bs\beta_n|\bs\mu_n,\bs\Sigma_n)$ with variational parameters $\mu_n$ and $\bs\Sigma_n$, the idea consists in learning a single parametric function $f_{\bs\theta}(\by_n)$ and considering a variational distribution of the form $q(\bs\beta_n|f_{\bs\theta}(\by_n))$. The function $f_{\bs\theta}(\by_n)$ maps the observed choices $\by_n$ of the $n$-th decision-maker (along with their corresponding covariates $\bX_n$ and alternative availability information $\textbf{a}_n$) to a set of variational parameters tailored to the taste parameters of that decision-maker. Naturally, this function needs to be sufficiently rich in order to capture the posterior accurately. Therefore, we propose to model this function using a deep neural network with parameters $\bs\theta$, whose architecture is depicted in Figure~\ref{fig:inference_net}. Essentially, we have introduced a generic \emph{inference network} that takes as input the observed data from a decision-maker, and outputs the approximate posterior distribution of her taste parameters $\bs\beta_n$. As a consequence, the number of variational parameters no longer grows with $N$, thereby ``amortizing'' its computational cost. The network parameters $\bs\theta$ can be estimated together with the remaining variational parameters $\bs\phi$ by optimizing the ELBO using stochastic backpropagation. Moreover, since we are using mini-batches, the use of an inference network $f_{\bs\theta}(\by_n)$ allows inference to generalize across batches and share statistical strength among different datapoints \citep{gershman2014amortized}. 

The proposed neural network architecture for the inference network is illustrated in Figure~\ref{fig:inference_net}. It takes as input the observations from the $n$-th decision-maker at different choice occasions (menus) $\{1,\dots,T\}$. The first layer consists of a 1-D convolutional layer that, for each menu $t$, takes as input the observed choice $y_{nt}$ (one-hot encoded), the covariates $\bX_{nt}$, and a binary vector $\textbf{a}_{nt}$ on length $J$ (number of alternatives) where $a_{nj} = 1$ indicates that the alternative $j$ is available. A stride of length $J+J*(K+L)+J$ is used for the convolution operation. The idea is that the convolution acts as a feature extractor from each menu $t$, thus building a lower-dimensional vector representation of the observed data $\{y_{nt}, \bX_{nt}, \textbf{a}_{nt}\},  \forall t \in {1,\dots,T}$. These lower-dimensional projections are then aggregated by a MaxPooling layer. The advantage of this approach is two-fold: (i) it allows the inference network to handle different numbers of menus $T$ for different decision-makers, and (ii) it significantly reduces the number of network parameters by sharing them across menus. The aggregated representation is then fed to a Batch Normalization layer\footnote{We have empirically found the use of Batch Normalization to significantly speed up training.} \citep{ioffe2015batch} and a fully-connected (FC) layer. Lastly, the inference network outputs a posterior approximation to the preference parameters $\bs\beta_n$ with the form of a multivariate normal, where the mean is determined by another FC layer. As for the covariance matrix, we make use of the Cholesky factorization. Concretely, we include two additional FC layers: one that uses a Softplus activation function (in order to ensure positivity) and outputs the diagonal elements of the Cholesky factor, and another with a linear activation function that outputs a lower triangular matrix corresponding to the remaining terms of the Cholesky factor, as depicted in Figure~\ref{fig:inference_net}. 
 
\subsection{Normalizing Flows}
\label{subsec:nfs}

So far, we have been assuming the variational distribution $q_{\bs\phi}(\bz|\by)$ to consist of well-known parametric distributional forms (\mbox{e.g. Gaussians}). However, for many datasets and MMNL models of interest (\mbox{e.g. with non-Gaussian mixing distributions}), approximating the true posterior distribution with a Gaussian using variational inference may be too strong of an assumption. In this section, we propose the use of normalizing flows (NFs) as a way to obtain complex density approximations to the true posterior. Normalizing flows \citep{rezende2015variational,papamakarios2019normalizing} provide a general mechanism for defining expressive probability distributions that only requires the specification of a (usually simple) base distribution and a series of bijective differentiable transformations (diffeomorphisms). Let $p(\textbf{u})$ be a probability distribution (\mbox{e.g.} isotropic Gaussian) and $T$ be an invertible transformation such that $T$ and $T^{-1}$ are both differentiable. Then, if $\textbf{u} = T^{-1}(\bz)$, the density of $\bz$ is given by 
\begin{align}
p(\bz) = p(\textbf{u}) \, | \det \textbf{J}_T(\textbf{u}) |^{-1},
\label{eq:change_volume}
\end{align}
where $\textbf{J}_T(\textbf{u}) = \frac{\partial T}{\partial \textbf{u}}$ is the Jacobian of the transformation. However, in order for us to be able to use NFs as variational approximations, the log determinant of the Jacobian in (\ref{eq:change_volume}) needs to be efficient to compute. As a consequence, several restrictions need to be imposed on $T$, and various authors have proposed different approaches for $T$ (see \cite{papamakarios2019normalizing} for a recent review). Although, at first sight, this assumption may be too restrictive, it turns out that due to the algebra of compositions inverses and Jacobian determinants, it is possible to compose multiple transformations in order to obtain increasingly flexible flows. Letting $T$ be a transformation composed of $K$ sub-transformation (\mbox{i.e.} $T = T_K \circ \cdots \circ T_1$), $p(\textbf{u})$ a base distribution, and $\textbf{z}_k$ be the variable that a sample $\bz_0 \sim p(\textbf{u})$ takes after the $k$-th transformation, then the log probability of the target distribution is given by
\begin{align}
\log p(\bz) = \log p(\textbf{u}) - \sum_{k=1}^K  \log | \det \textbf{J}_{T_k}(\bz_{k-1}) |.
\end{align}

In this paper, we consider Sylvester normalizing flows \citep{berg2018sylvester}, which rely on a transformation of the form
\begin{align}
\bz_k = \bz_{k-1} \textbf{Q} \textbf{R} \, h(\tilde{\textbf{R}} \textbf{Q}^\transpose \bz_{k-1} + \textbf{b}), 
\label{eq:sylvester}
\end{align}
where $h$ is suitable smooth activation function, $\textbf{R}$ and $\tilde{\textbf{R}}$ are upper triangular $M \times M$ matrices and $\textbf{Q} = (\textbf{q}_1,\dots,\textbf{q}_M)$, with columns $\textbf{q}_m \in \mathbb{R}^D$ forming an orthonormal set of vectors. Under these conditions, the transformation in (\ref{eq:sylvester}) is invertible and the determinant of the Jacobian $\textbf{J}$ can be computed in $\mathcal{O}(M)$ (see \cite{berg2018sylvester} for details). This transformation that resembles a multi-layer FC neural network therefore provides an expressive building block for constructing arbitrarily complex normalizing flows. 

\section{Experiments}
\label{sec:experiments}

In order to evaluate the performance of the proposed approach, both in terms of scalability and estimation accuracy, we performed a simulation study. Namely, we pre-specified the values of $\bs\alpha$, $\bs\zeta$ and $\bs\Omega$, and then generated artificial datasets of arbitrary sizes $N$ and $T$ using the data generating process described in Section~\ref{sec:mmnl}: for each decision-maker $n$ we sampled random taste parameters $\bs\beta_n \sim \N(\bs\zeta,\bs\Omega)$, and for each choice occasion $t$ we sampled $y_{nt} \sim \mbox{MNL}(\bs\alpha, \bs\beta_n, \bX_{nt})$. The covariates were assumed to be uniformly distributed $\bx_{ntj} \sim \mathcal{U}(0,1)$. 

The proposed approach was implemented in Pyro \citep{bingham2019pyro} and PyTorch. We consider 3 different implementation variants: 
\begin{itemize}
\item a pure Stochastic Variational Inference (``SVI'') approach \citep{hoffman2013stochastic}, where the number of variational parameters grows with $N$;
\item a simpler Amortized Variational Inference (``AVI'') approach, where the inference network only outputs the posterior mean of the approximation ($\bs\mu_n$ in Figure~\ref{fig:inference_net}), while the covariance of the posterior approximation is assumed to be a shared learnable parameter represented by its Cholesky factorization;
\item the full Amortized Variational Inference approach, using the inference network from Figure~\ref{fig:inference_net}, which we refer as ``AVI2''.
\end{itemize}

\subsection{Scalability}

We compare the performance of these 3 variants with MSLE and Gibbs sampling, based on the efficient multi-thread implementations provided by \cite{bansal2020bayesian}. All experiments were performed on a single machine with an 8-core CPU @ 2.40GHz, 64GB of RAM and a GeForce GTX 1080 TI GPU. Table~\ref{table:results} shows the obtained results for various values of $N$ and $T$. The number of fixed and random taste parameters were kept fixed at $L=3$ and $K=5$, respectively. The number of alternatives used was $J=5$. In order to keep the analysis concise, we focus on the more interesting scenario where the correlation between random taste parameters is high. All experiments were repeated 30 times and the average results (and standard deviations) are reported. 

\begin{table}[t!]
\caption{Results obtained for the different approaches.}
\label{table:results}
\centering
\small
\begin{tabular}{lll llll}
\toprule
\multicolumn{7}{l}{$N = 500$; $T = 5$; $J = 5$; $L = 3$; $K = 5$; Batch Size = 500}\\ 
Method & Runtime (s) & Sim. Loglik. & RMSE $\bs\alpha$ & RMSE $\bs\zeta$ & RMSE $\bs\beta_n$ & RMSE $\bs\Omega$  \\
\midrule
MSLE & 176 ($\pm$24) & -3475 ($\pm$34) & 0.081 ($\pm$0.034) & 0.094 ($\pm$0.033) & 0.785 ($\pm$0.025) & 0.240 ($\pm$0.063)\\
Gibbs & 227 ($\pm$6) & -3477 ($\pm$34) & 0.080 ($\pm$0.035) & 0.095 ($\pm$0.033) & 0.777 ($\pm$0.024) & 0.213 ($\pm$0.062)\\
SVI-LKJ & 125 ($\pm$17) & -3483 ($\pm$34) & 0.078 ($\pm$0.032) & 0.093 ($\pm$0.034) & 0.790 ($\pm$0.025) & 0.267 ($\pm$0.049)\\
AVI-LKJ & 127 ($\pm$21) & -3482 ($\pm$34) & 0.078 ($\pm$0.034) & 0.093 ($\pm$0.033) & 0.792 ($\pm$0.024) & 0.259 ($\pm$0.050)\\
AVI2-LKJ & 176 ($\pm$34) & -3484 ($\pm$34) & 0.089 ($\pm$0.031) & 0.097 ($\pm$0.032) & 0.799 ($\pm$0.025) & 0.285 ($\pm$0.052)\\
\midrule
\multicolumn{7}{l}{$N = 2000$; $T = 10$; $J = 5$; $L = 3$; $K = 5$; Batch Size = 2000}\\ 
Method & Runtime (s) & Sim. Loglik. & RMSE $\bs\alpha$ & RMSE $\bs\zeta$ & RMSE $\bs\beta_n$ & RMSE $\bs\Omega$  \\
\midrule
MSLE & 2537 ($\pm$713) & -27595 ($\pm$106) & 0.030 ($\pm$0.015) & 0.036 ($\pm$0.013) & 0.659 ($\pm$0.006) & 0.081 ($\pm$0.022)\\
Gibbs & 1257 ($\pm$713) & -27596 ($\pm$106) & 0.029 ($\pm$0.015) & 0.036 ($\pm$0.013) & 0.657 ($\pm$0.007) & 0.075 ($\pm$0.024)\\
SVI-LKJ & 97 ($\pm$13) & -27614 ($\pm$105) & 0.030 ($\pm$0.015) & 0.040 ($\pm$0.013) & 0.660 ($\pm$0.007) & 0.115 ($\pm$0.027)\\
AVI-LKJ & 90 ($\pm$12) & -27612 ($\pm$105) & 0.030 ($\pm$0.014) & 0.037 ($\pm$0.011) & 0.659 ($\pm$0.007) & 0.096 ($\pm$0.025)\\
AVI2-LKJ & 147 ($\pm$22) & -27613 ($\pm$105) & 0.031 ($\pm$0.015) & 0.038 ($\pm$0.014) & 0.660 ($\pm$0.007) & 0.088 ($\pm$0.023)\\
\midrule
\multicolumn{7}{l}{$N = 5000$; $T = 10$; $J = 5$; $L = 3$; $K = 5$; Batch Size = 2000}\\ 
Method & Runtime (s) & Sim. Loglik. & RMSE $\bs\alpha$ & RMSE $\bs\zeta$ & RMSE $\bs\beta_n$ & RMSE $\bs\Omega$  \\
\midrule
MSLE & 9003 ($\pm$22) & -68907 ($\pm$135) & 0.017 ($\pm$0.008) & 0.022 ($\pm$0.007) & 0.658 ($\pm$0.005) & 0.050 ($\pm$0.012)\\
Gibbs & 4910 ($\pm$12) & -68906 ($\pm$137) & 0.017 ($\pm$0.009) & 0.022 ($\pm$0.006) & 0.656 ($\pm$0.005) & 0.047 ($\pm$0.011)\\
SVI-LKJ & 118 ($\pm$14) & -68965 ($\pm$137) & 0.021 ($\pm$0.010) & 0.038 ($\pm$0.013) & 0.664 ($\pm$0.006) & 0.152 ($\pm$0.034)\\ 
AVI-LKJ & 114 ($\pm$18) & -68943 ($\pm$133) & 0.020 ($\pm$0.008) & 0.026 ($\pm$0.008) & 0.662 ($\pm$0.005) & 0.095 ($\pm$0.019)\\
AVI2-LKJ & 155 ($\pm$23) & -68940 ($\pm$133) & 0.019 ($\pm$0.010) & 0.028 ($\pm$0.010) & 0.661 ($\pm$0.005) & 0.079 ($\pm$0.020)\\
\midrule
\midrule
\multicolumn{7}{l}{$N = 10000$; $T = 10$; $J = 5$; $L = 3$; $K = 5$; Batch Size = 2000}\\ 
Method & Runtime (s) & Sim. Loglik. & RMSE $\bs\alpha$ & RMSE $\bs\zeta$ & RMSE $\bs\beta_n$ & RMSE $\bs\Omega$  \\
\midrule
SVI-LKJ & 158 ($\pm$21) & -138017 ($\pm$171) & 0.022 ($\pm$0.009) & 0.057 ($\pm$0.020) & 0.671 ($\pm$0.007) & 0.214 ($\pm$0.038)\\
AVI-LKJ & 142 ($\pm$24) & -137930 ($\pm$158) & 0.016 ($\pm$0.009) & 0.024 ($\pm$0.007) & 0.663 ($\pm$0.003) & 0.087 ($\pm$0.020)\\
AVI2-LKJ & 156 ($\pm$25) & -137931 ($\pm$160) & 0.017 ($\pm$0.007) & 0.022 ($\pm$0.009) & 0.663 ($\pm$0.003) & 0.075 ($\pm$0.021)\\
\midrule
\multicolumn{7}{l}{$N = 50000$; $T = 10$; $J = 5$; $L = 3$; $K = 5$; Batch Size = 2000}\\ 
Method & Runtime (s) & Sim. Loglik. & RMSE $\bs\alpha$ & RMSE $\bs\zeta$ & RMSE $\bs\beta_n$ & RMSE $\bs\Omega$  \\
\midrule
SVI-LKJ & 3502 ($\pm$373) & -709140 ($\pm$378) & 0.062 ($\pm$0.005) & 0.052 ($\pm$0.009) & 0.994 ($\pm$0.003) & 0.834 ($\pm$0.000)\\
AVI-LKJ & 634 ($\pm$86) & -689680 ($\pm$465) & 0.015 ($\pm$0.007) & 0.022 ($\pm$0.008) & 0.670 ($\pm$0.002) & 0.109 ($\pm$0.012)\\
AVI2-LKJ & 648 ($\pm$115) & -689638 ($\pm$459) & 0.014 ($\pm$0.006) & 0.021 ($\pm$0.006) & 0.669 ($\pm$0.002) & 0.093 ($\pm$0.012)\\
\bottomrule
\end{tabular}
\end{table}

From the results in Table~\ref{table:results}, it can observed that for a small number of decision-makers ($N=500$) the computational times for the different estimation methods are relatively similar, and so are the estimation results (i.e. simulated likelihoods and RMSEs for $\bs\alpha$, $\bs\zeta$, $\bs\beta_n$ and $\bs\Omega$). However, as we increase $N$, the differences in runtime quickly become dramatically significant. It can observed that for a relatively small dataset ($N=2000$ and $T=10$), MSLE and Gibbs sampling take an average of 42 and 21 minutes respectively (even though they are taking advantage of the 8 CPU cores of the machine used in the experiments), while ``SVI-LKJ'' and ``AVI-LKJ'' are able to achieve similar estimation errors in less than 2 minutes. The ``AVI2-LKJ'' variant takes slightly longer due to the extra number of parameters in the inference network. 

\begin{figure}[t!]
\begin{center}
\includegraphics[width=0.9\linewidth,trim=0 0 0 0, clip]{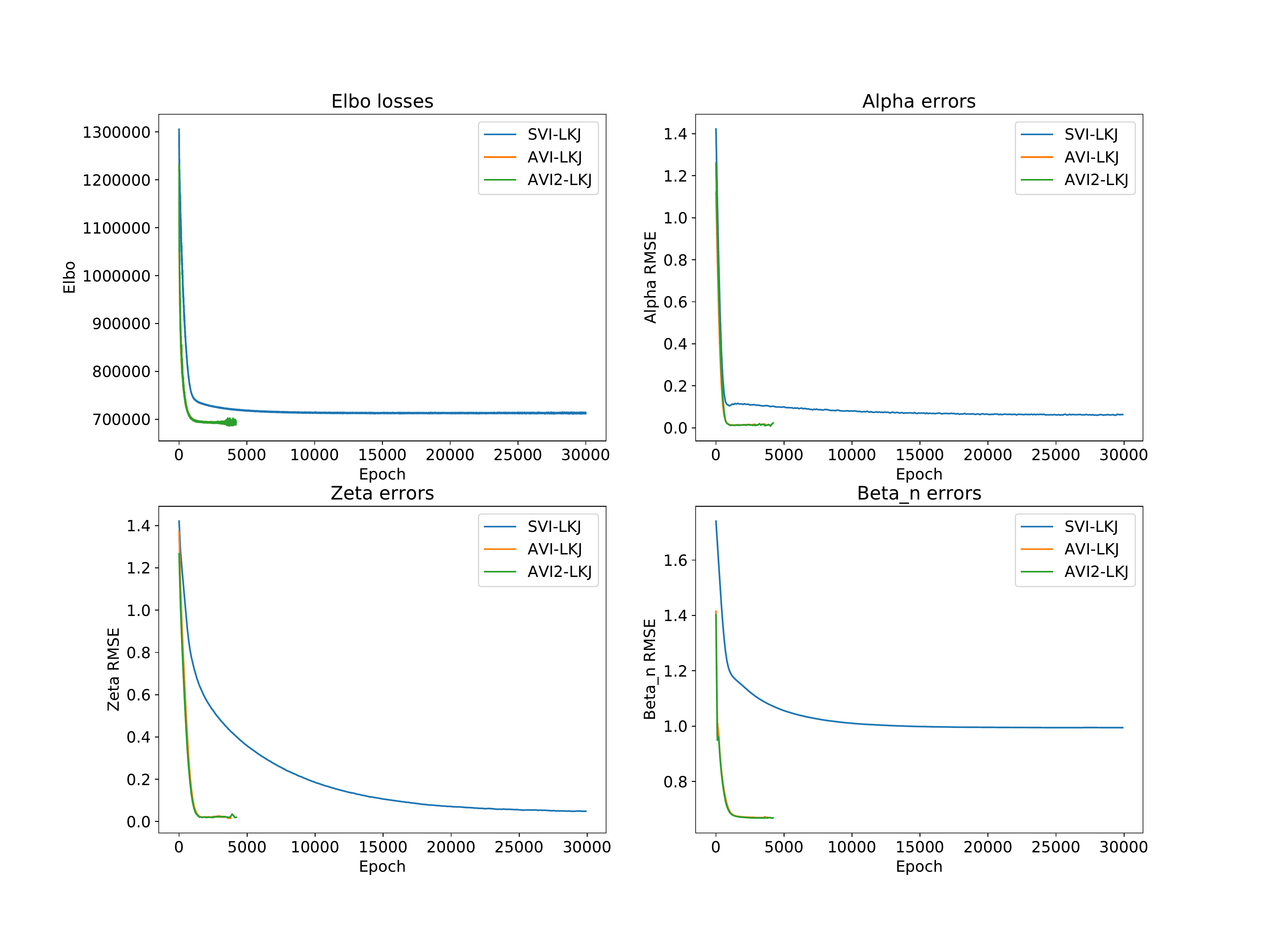}
\caption{Convergence plot comparing ``SVI-LKJ'' (blue), ``AVI-LKJ'' (orange) and ``AVI2-LKJ'' (green) according to different metrics. The lines for ``AVI-LKJ'' and ``AVI2-LKJ'' are difficult to distinguish because they are on top of each other.}
\label{fig:convergence_plot}
\end{center}
\end{figure}

Up to $N=2000$, the differences in the computational times observed can be attributed to the well-known efficiency of variational inference methods \citep{braun2010variational,bansal2020bayesian} and to the use of GPU-accelarated algebra from PyTorch. We have yet to fully witness the advantages of mini-batching and amortization. In order to do so, we considered larger datasets with sizes $N=5000$, $N=10000$ and $N=50000$. As the results from Table~\ref{table:results} show, while for $N=5000$ we were still able to run MSLE and Gibbs sampling (although they took an average of 150 and 82 minutes to run, respectively), for larger $N$ our 64GB RAM proved insufficient. On the contrary, for $N=5000$, our variational inference implementations were able to achieve similar results in approximately 2 minutes. Interestingly, while for $N=5000$ we don't observe a significant difference between SVI and AVI, for larger $N$, ``SVI-LKJ'' starts to exhibit estimation problems due to the fact that the number of local variational parameters grows with the number of decision-makers $N$ - a problem that is aggravated by the use of mini-batching. On the other hand, the approaches based on Amortized Variational Inference don't suffer from this issue, and continue to perform well even for extremely large datasets. Indeed, we can verify that ``AVI-LKJ'' and ``AVI2-LKJ'' are able to perform approximate Bayesian inference in a very large dataset with 50000 decision-makers in approximately 10 minutes. In order to understand this issue better, Figure~\ref{fig:convergence_plot} shows the convergence of the different VI variants according to different metrics. While ``SVI-LKJ'' can take many epochs (\mbox{i.e.} passes through the dataset) in order to start achieving reasonable error values, both ``AVI-LKJ'' and ``AVI2-LKJ'' converge to the solution in a very small number of epochs due to their ability to ``amortize'' the cost of inference by sharing statistical strength between mini-batches. 

\begin{figure}[t!]
\begin{center}
\includegraphics[width=0.4\linewidth,trim=0 0 0 0, clip]{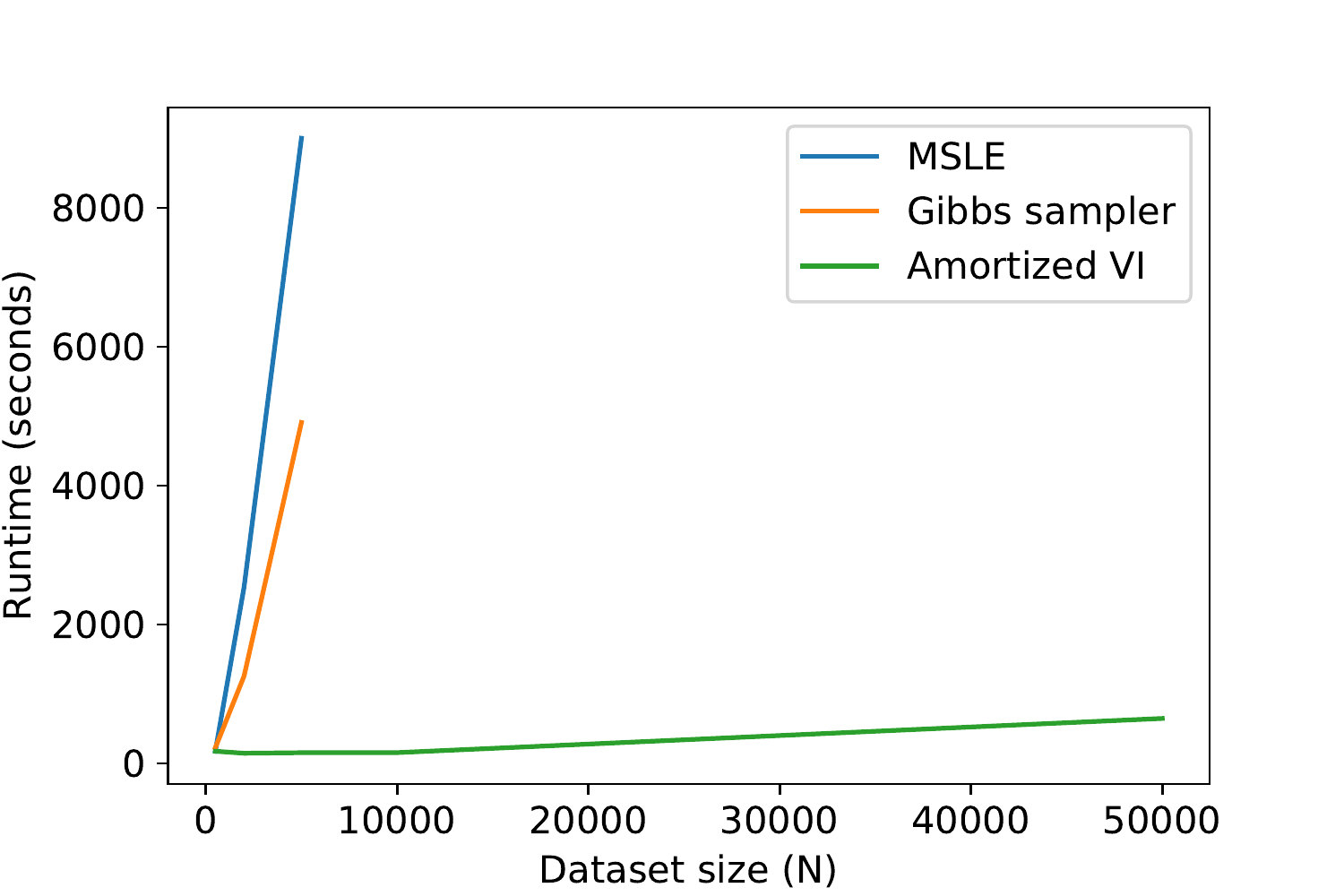}
\caption{Scalability plot comparing Amortized Variational Inference, MSLE and Gibbs sampling.}
\label{fig:scalability_plot}
\end{center}
\end{figure}

Lastly, in order to provide a better representation of the scalability of the proposed Amortized Variational Inference approach when compared to MSLE and Gibbs sampling, Figure~\ref{fig:scalability_plot} plots the runtime as a function of $N$. Analysing the trend in this plot illustrates well the scalability of the proposed Amortized Variational Inference approach and how it differs dramatically from the scalability of MSLE and Gibbs sampling. 

\subsection{Out-of-sample generalization}

An important advantage of the proposed Amortized Variational Inference approach is that, upon convergence, not only are we provided with approximate posterior distributions for all the latent variables in the MMNL model, but we also have access to a trained inference network $f_{\bs\theta}(\cdot)$. In other words, we have at our disposal a neural network that given a set of observed choices $\by_{n^*}$ and corresponding covariates $\bX_{n^*}$ for a new (unseen) decision-maker $n^*$, we can approximate the posterior distribution of her random taste parameters with a single forward pass through the inference network $f_{\bs\theta}(\cdot)$. In order to evaluate that capacity of the inference network to generalize to out-of-sample choice data, we generated an additional testset of the same size as the trainset and using the same data-generating process described in Section~\ref{sec:mmnl}. We then ran the learned inference network on this testset and compared its ability to approximate the true random taste parameters with the quality of the estimates on the trainset. Table~\ref{table:results_cf} shows the obtained results for different values of $N$. Unsurprisingly, the larger the size of the trainset, the closer the gap between train and testset RMSE is in terms of $\bs\beta_n$ estimation, which in turn results in more similar accuracies and loglikelihoods for the two sets. Although these are preliminary results with artificially-generated data, and more experiments are required with real choice datasets, they are very encouraging for example for real-time applications \citep{cottrill2013future} and online learning \citep{danaf2019online}. Moreover, having an inference network that generalizes well to out-of-sample data opens up the possibility to efficiently explore counterfactuals of the form ``what would be the taste parameter of a decision-maker $n$, or the difference in her value-of-time, if the she had chosen alternative $j$ in choice occasion $t$?''. 


\begin{table}[t!]
\caption{Results for the generalization ability of the inference network ($T = 10$; $J = 5$; $L = 3$; $K = 5$).}
\label{table:results_cf}
\centering
\small
\begin{tabular}{l | ll | ll | ll}
\toprule
 & \multicolumn{2}{c}{Loglikelihood} & \multicolumn{2}{c}{Accuracy} & \multicolumn{2}{c}{RMSE $\bs\beta_n$}\\
N & Train & Test & Train & Test & Train & Test\\
\midrule
500 & 6139 ($\pm$79) & 6802 ($\pm$49) & 0.502 ($\pm$0.009) & 0.432 ($\pm$0.008) & 0.670 ($\pm$0.014) & 0.883 ($\pm$0.018)\\  
2000 & 24785 ($\pm$152) & 26044 ($\pm$106) & 0.496 ($\pm$0.005) & 0.462 ($\pm$0.004) & 0.661 ($\pm$0.007) & 0.767 ($\pm$0.011)\\  
10000 & 125760 ($\pm$289) & 127328 ($\pm$282) & 0.485 ($\pm$0.002) & 0.475 ($\pm$0.002) & 0.659 ($\pm$0.003) & 0.694 ($\pm$0.004)\\  
\bottomrule
\end{tabular}
\end{table}

\subsection{Normalizing flows}

We will now investigate the use of normalizing flows as a way of obtaining a richer family of variational distributions $q$. In order to do so, we altered the generative process of the MMNL model from Section~\ref{sec:mmnl}, such that $\bs\zeta \sim \mbox{LogNormal}(0,1)$. Three different stochastic variational inference approaches with different variational approximation $q(\bs\zeta)$ were then used to perform inference in this model: (i) Gaussian approximation, (ii) LogNormal approximation and (iii) a Sylvester normalizing flow with a isotropic K-dimensional Gaussian as the base distribution. The parameters of the base distribution of the normalizing flow were also learnable. Table~\ref{table:results_nfs} shows the obtained results for different numbers of decision-makers $N$ and choice occasions $T$. The latter suggest that normalizing flows can indeed be a valuable technique for approximating more complex posteriors, therefore leading to higher loglikelihood and simulated loglikelihood scores, with little computational overhead over simpler parametric assumptions. These results are in line with previous works, that evidence the ability of normalizing flows to capture complex posteriors \citep{rezende2015variational,berg2018sylvester}, including multi-modal ones \citep{papamakarios2019normalizing}, thereby providing a promising direction for accurately approximating rich posterior distributions in discrete choice models (which is often pointed out as a weakness of VI in contrast with MCMC methods), without significantly compromising computational efficiency. 

\begin{table}[t!]
\caption{Results obtained by Sylvester normalizing flows (NFs), in comparison with two baseline approximations.}
\label{table:results_nfs}
\centering
\small
\begin{tabular}{llll}
\toprule
\multicolumn{4}{l}{$N = 500$; $T = 5$; $J = 5$; $L = 3$; $K = 5$; Batch Size = 500}\\ 
Method & Runtime (s) & Loglikelihood  & Sim. Loglik.\\
\midrule
SVI-LKJ (LogNormal) & 223 ($\pm$3) & -2981 ($\pm$49) & -3034 ($\pm$45)\\  
SVI-LKJ (Normal) & 222 ($\pm$2) & -2968 ($\pm$50) & -3025 ($\pm$46)\\ 
SVI-LKJ (NFs) & 272 ($\pm$3) & -2961 ($\pm$50) & -3020 ($\pm$46)\\ 
\midrule
\multicolumn{4}{l}{$N = 500$; $T = 10$; $J = 5$; $L = 3$; $K = 5$; Batch Size = 500}\\ 
Method & Runtime (s) & Loglikelihood & Sim. Loglik.\\
\midrule
SVI-LKJ (LogNormal) & 223 ($\pm$1) & -6156 ($\pm$76) & -6239 ($\pm$73)\\  
SVI-LKJ (Normal) & 222 ($\pm$2) & -6154 ($\pm$76) & -6238 ($\pm$72)\\ 
SVI-LKJ (NFs) & 268 ($\pm$3) & -6137 ($\pm$75) & -6227 ($\pm$72)\\ 
\midrule
\multicolumn{4}{l}{$N = 2000$; $T = 5$; $J = 5$; $L = 3$; $K = 5$; Batch Size = 2000}\\ 
Method & Runtime (s) & Loglikelihood & Sim. Loglik.\\
\midrule
SVI-LKJ (LogNormal) & 232 ($\pm$3) & -12162 ($\pm$100) & -12319 ($\pm$94)\\  
SVI-LKJ (Normal) & 232 ($\pm$2) & -12115 ($\pm$100) & -12283 ($\pm$94)\\ 
SVI-LKJ (NFs) & 279 ($\pm$3) & -12086 ($\pm$103) & -12261 ($\pm$95)\\ 
\midrule
\multicolumn{4}{l}{$N = 2000$; $T = 10$; $J = 5$; $L = 3$; $K = 5$; Batch Size = 2000}\\ 
Method & Runtime (s) & Loglikelihood & Sim. Loglik.\\
\midrule
SVI-LKJ (LogNormal) & 235 ($\pm$2) & -24865 ($\pm$154) & -25150 ($\pm$146)\\  
SVI-LKJ (Normal) & 232 ($\pm$2) & -24856 ($\pm$157) & -25144 ($\pm$148)\\ 
SVI-LKJ (NFs) & 278 ($\pm$2) & -24789 ($\pm$154) & -25093 ($\pm$145)\\ 
\bottomrule
\end{tabular}
\end{table}

\section{Conclusion}
\label{sec:conclusion}

This paper proposed an Amortized Variational Inference approach for effectively scaling Bayesian inference in MMNL models to very large datasets. Through an empirical simulation study, we demonstrated that its combination of an expressive (and yet compact) inference network, stochastic backpropagation, automatic differentiation and GPU-accelerated computation (using easily-accessible inexpensive hardware), the proposed approach is able to achieve computational speedups of multiple orders of magnitude over traditional MSLE and MCMC approaches for large datasets without compromising estimation accuracy. Therefore, this effectively opens up the way for a new wave of scalable, fully-Bayesian, VI-powered MMNL toolboxes whose efficiency would be highly appreciated by researchers and practitioners. Moreover, the use of Monte Carlo gradient estimation techniques and probabilistic programming languages such as Pyro \citep{bingham2019pyro}, makes the entire modelling framework extremely flexible, thus speeding up the development and testing of new modelling approaches. Lastly, our preliminary results with the use of normalizing flows for constructing flexible variational distributions that can accurately approximate complex posteriors were encouraging. Normalizing flows is currently a very active area of research with new approaches being constantly proposed, and we plan to keep exploring their potential in discrete choice modelling as part of our future work, which will also consider real-world datasets and the use of the inference network for counterfactual reasoning.


\bibliography{mxl-amortized-vi}

\end{document}